\documentclass[journal]{IEEEtran}

\usepackage{mathtools}
\usepackage[makeroom]{cancel}
\usepackage{amssymb}
\usepackage{graphicx}
\usepackage[english]{babel}
\usepackage{thmtools}
\usepackage{color}
\usepackage{caption}
\usepackage[framemethod=tikz]{mdframed}
\usepackage{lipsum}
\usepackage{comment}
\usepackage{array}
\usepackage{changepage}
\usepackage{bm}
\usepackage{amsbsy}
\usepackage{makecell}
\usepackage{multirow}
\usepackage{amsmath}
\usepackage[margin=1in]{geometry}
\usepackage{booktabs,threeparttable}
\usepackage{algorithm}
\usepackage{longtable}
\usepackage{verbatimbox}
\usepackage{xtab,afterpage}
\usepackage{pgffor}
\usepackage{tikz}
\usepackage{psfrag}
\usepackage{subcaption}

\newcolumntype{P}[1]{>{\centering\arraybackslash}p{#1}}
\newcommand{\bottomstrutH}[1]{\rule[-#1]{0pt}{#1}}
\newcommand{\topstrutH}[1]{\rule{0pt}{#1}}
\newcommand{\dividerline}[2]{\bottomstrutH{#1}\\\hline\topstrutH{#2}}
	\DeclareCaptionStyle{mystylefigures}%
	[margin=5mm,justification=centering]%
	{font=footnotesize,margin={5mm,3mm}, labelsep = period}
\DeclareCaptionStyle{mystyletables}%
[margin=5mm,justification=centering]%
{font=footnotesize,labelfont=sc, textfont=sc, margin={8mm,5mm}, labelsep = period}
\ifCLASSINFOpdf

\else

\fi

\newcommand{\nn}{\nonumber}
\newcommand{\mc}{\mathcal}

\newcommand{\beq}{\begin{equation}}
\newcommand{\eeq}{\end{equation}}

\newcommand{\st}{\text{s.t.}}

\IEEEoverridecommandlockouts

\usepackage{ifthen}
\newboolean{showcomments}
\setboolean{showcomments}{true}
\newcommand{\yue}[1]{\ifthenelse{\boolean{showcomments}}
{ \textcolor{red}{(Yue says:  #1)}}{}}

\newcommand{\hossein}[1]{\ifthenelse{\boolean{showcomments}}
{ \textcolor{blue}{(Hossein says:  #1)}}{}}

\newcommand{\BEAS}{\begin{eqnarray*}}
\newcommand{\EEAS}{\end{eqnarray*}}
\newcommand{\BEQ}{\begin{equation}}
\newcommand{\EEQ}{\end{equation}}
\newcommand{\BIT}{\begin{itemize}}
\newcommand{\EIT}{\end{itemize}}

\newtheorem{theorem}{Theorem}

\newtheorem{RK}{Remark}

\usepackage{psfrag}

\usepackage{nomencl}

\makenomenclature

\usepackage{makeidx}
\makeindex

\usepackage{tikz}
\usetikzlibrary{shapes,backgrounds,calc}

\makeatletter
\tikzset{circle split part fill/.style  args={#1,#2}{%
 alias=tmp@name, 
  postaction={%
    insert path={
     \pgfextra{%
     \pgfpointdiff{\pgfpointanchor{\pgf@node@name}{center}}%
                  {\pgfpointanchor{\pgf@node@name}{east}}%
     \pgfmathsetmacro\insiderad{\pgf@x}
      \fill[#1] (\pgf@node@name.base) ([xshift=-\pgflinewidth]\pgf@node@name.east) arc
                          (0:180:\insiderad-\pgflinewidth)--cycle;
      \fill[#2] (\pgf@node@name.base) ([xshift=\pgflinewidth]\pgf@node@name.west)  arc
                           (180:360:\insiderad-\pgflinewidth)--cycle;            
         }}}}}
 \makeatother

\usepackage{pgffor}

\usepackage{algorithmic}
\usepackage{arydshln}

\begin{document}
\title{{On the Equilibria and Efficiency of Electricity Markets with Renewable Power Producers and Congestion Constraints}}
\author{{Hossein Khazaei, X. Andy Sun, and Yue Zhao
}
\thanks{The authors are listed alphabetically.}
} 
\maketitle

\begin{abstract}
With increasing renewable penetration in power systems, a prominent challenge in the efficient and reliable power system operation is handling the uncertainties inherent in the renewable {generation}. In this paper, we propose a simple two-settlement market mechanism in which renewable power producers (RPPs) participate so that a) the independent system operator (ISO) does not need to consider the uncertainties of the renewables in its economic dispatch, and yet b) the market equilibrium is shown to approach social efficiency as if the ISO solves a stochastic optimization problem taking into account all the uncertainties. In showing this result, a key innovation is a new approach of efficiently computing the Nash equilibrium (NE) among the strategic RPPs in congestion-constrained power networks. 
In particular, the proposed approach decouples finding an NE into searching over congestion patterns and computing an NE candidate assuming a congestion pattern. As such, the computational complexity of finding an NE grows only polynomially with the number of RPPs in the market. We demonstrate our results in the IEEE 14-bus system and show that the NE approaches social efficiency as the number of RPPs grows. 

%
\end{abstract}

\section{Introduction}

Renewable {energy} plays a central role in achieving a sustainable energy future. However, renewable {generation} such as wind and solar power are inherently non-dispatchable, and highly uncertain and variable. As a result, integrating renewable {energy} into power systems 
raises significant {challenges on} reliability and efficiency {for power system operations} \cite{Pinson2013}. 
In particular, in {operating an electricity} market with renewable {generation}, how to take into account their uncertainties in order to achieve optimal system operation is a major issue and an active area of research. 
%

\subsection{Background and related work}
The primary goal of an ISO who runs {an electricity} market is to maximize the social welfare (or, equivalently, minimizing the social cost) while assuring secure power system operation. Ideally, achieving such goals in the presence of the uncertain renewables can be formulated as a (potentially multi-stage and multi-period) \emph{stochastic optimization} problem \cite{ConejoBook,shapiro2009lectures, Hobbs2012}. 

{Recent progress has substantially improved the scalability of multistage stochastic programs (see e.g. \cite{ZouSDDiP2019,ZouUC2018}). However, significant challenges still remain for handling mixed integer decisions and contingency constraints. Another more subtle but fundamental challenge} is that, for a stochastic program to be employed in an electricty market, a) the probability distributions of all underlying stochastic processes must be known to and agreed upon by all market participants, including the ISO, and b) a particular sample average approximation of the stochastic processes and the associated set of scenarios also must be accepted to all market participants. It is a very challenging task to accurately a) estimate and b) form a consensus of these distributions when critical information of market participants remain private in a market setting. 




In addition to {challenges in modeling and computation of} the optimal system operation problem, another major concern of the ISOs' is the \emph{strategic behaviors} of market participants. Indeed, an ISO typically depends on market participants to provide information on their resources such as generation cost/willingness to generate, capacity, and other constraints. Assuming that the participants provide their true information, an ISO can then solve for the optimal system operation that minimizes the social cost. However, market participants do not necessarily behave truthfully, but rather {may behave} strategically by providing information that benefit themselves. {Indeed, such strategic behavior in electricity market has been a subject of intense study, 
in particular, via the analysis of Nash equilibria (NE) in the market clearing game played between the market participants and the ISO}. 

Several approaches have been proposed to compute NE in {electricity} markets, albeit not yet considering uncertain generation from RPPs. One is to solve equilibrium problems with equilibrium constraints (EPEC) \cite{ConejoBook, Hu07}. 
This approach is however not without limitations, as its high computational complexity can become a major bottleneck as the number of strategic players increases. Another approach is to solve for supply function equilibrium (SFE) \cite{anderson2002using, anderson2008finding, johari2011, NaLi15, Xu16}. It is however technically challenging for this approach to generalize to congested power networks, {although} recent progress has been made {in this direction} \cite{lin2017structural}. {The electricity market model has also been simplified to Cournot competition. Computation and analysis of NE in such settings} have been studied in \cite{allaz1993cournot, yao2008modeling, cai2017role}.

Considering RPPs with uncertain generation as participants in {a two-settlement market}, NE among RPPs has been studied only for limited settings. 
Most of the work assume {price-taking} RPPs and do not consider network congestion constraints. The key issue here boils down to payoff allocation among an aggregation of RPPs \cite{Nayyar13, Bitar11, Harirchi14, Zhao2016}. A recent work \cite{Zhao19} puts forward a simple payoff allocation mechanism that achieves a set of {desirable} properties, including achieving the maximum social welfare at the unique NE among the RPPs. 
Generalizing the price-taking assumption to price-making, albeit still not considering network constraints, another recent work proposes a market mechanism for RPPs with which the NE enjoys an analytical form and is proved to converge to social efficiency as the number of RPPs grows \cite{ZhaoPES18}. 

\subsection{Contributions of this work}
This paper a) proposes a two-settlement market mechanism for RPPs that takes full account of the power network congestion constraints, b) develops a new method for efficiently computing the NE of an electricity market, in this work specifically applied to our proposed mechanism, and c) demonstrates that the proposed market mechanism for RPPs leads to NE that approach social efficiency as the number RPPs grows.  



The proposed market mechanism is in fact quite intuitive, and not far from the current {practice of the ISOs}, but 
simpler. Specifically, a) in the day-ahead (DA) market, 
each RPP only submits a single number {of its production level, a ``commitment''}, to the ISO; b) the ISO treats the RPPs' commitments as \emph{firm}, performs the DA optimal dispatch without considering any uncertainty, and pays the RPPs using the DA locational marginal prices (LMPs); c) in the real-time (RT) market, the RPPs generation are realized; 
and d) the ISO performs the RT optimal dispatch to resolve all the power imbalances, and pays and/or charges the RPPs using the RT LMPs and according to their realized generation's deviations from the DA commitments. 
As such, in the proposed market mechanism, the ISO is freed from {the burden of modeling renewable} uncertainties, and {is only concerned with solving a deterministic market clearing problem}. 

To be able to evaluate the outcome of the proposed mechanism under strategic behaviors of the RPPs, we develop a new method for computing the NE of the {two-settlement} market. The key {idea} lies in the fact that, if a congestion pattern is given, computing the NE is not much more {complicated} than the uncongested case. Accordingly, we decouple the search for NE into a) search over congestion patterns, and b) computing the NE candidates given congestion patterns. As such, {we can show that} the computational complexity of the proposed method for computing NE scales only cubically with the number of RPPs. {Furthermore}, recent results from \cite{Tong17, Javadi17, Roald18, Roald19} as well as {the} conventional wisdom in practice both suggest that the number of potential congestion patterns in a real-world power system is relatively small. {Therefore, the proposed method can be quite efficient in real-world systems}. In {computation experiments}, we demonstrate {the proposed} mechanism and {the computed} NE in a congested IEEE 14-bus system. We show that the NE approaches social efficiency as the number of RPPs grows. The intuition is that the competition among the RPPs pushes the NE toward social efficiency, even though the ISO does not consider any uncertainty in its DA and RT dispatches in the proposed mechanism. 


\section{System model and proposed mechanism} \label{sec::SysMdl}
We consider a two-settlement market consisting of a Day-Ahead (DA) market (forward) and a Real-Time (RT) market (spot) run by an ISO. 
The participants of this two-settlement market include:
\begin{itemize}
    \item Renewable Power Producers (RPPs).
	\item DA conventional generators that can be dispatched in the DA market.
	\item RT conventional generators that can be dispatched in the RT market.
	\item Loads. 
\end{itemize}
Note that the set of DA and RT conventional generators can have an arbitrary overlap. In other words, the same generator, if capable, can choose to participate in both the DA and RT markets. 
In this study, our focus is on the \emph{strategic behaviors of RPPs}. 
As such, we assume that a) the conventional generators are truthful in submitting to the ISO their costs, capacities, etc., and b) the loads are inelastic and known. 


\subsection{Proposed mechanism with price-making RPPs}  \label{subsec::TwoSttlmntMrkt}
Motivated by the near-zero variable cost of renewable generation, 
we propose a two-settlement market where both conventional generators and RPPs participate in as follows: 
\begin{enumerate}
	\item In the DA market:
	\begin{enumerate}
		\item Each DA conventional generator submits its bidding curve to the ISO.
		\item Each RPP $k$ submits a \emph{firm commitment} $c_k$ for its power delivery at RT. 
		\item Upon receiving these information, the ISO considers the RPPs' DA commitments as firm, i.e., negative loads, and performs an optimal dispatch of the DA conventional generators to meet all the loads. 
		\item The ISO pays each RPP using the DA LMPs computed from the optimal DA dispatch, in the amount of $\lambda_{\Omega_{k}}^D   c_k$. 
	\end{enumerate}
	\item In the RT market,
	\begin{enumerate}
		\item Each RT conventional generator submits its bidding curve to the ISO. 
		\item Each RPP's actual generation $x_k$ is realized.
		\item The ISO 
		resolves all the deviations between the RPPs' DA commitments and their realized generation by optimally dispatching the RT conventional generators. 
		\item The ISO pays each RPP using the RT LMPs computed from the optimal RT dispatch, in the (possibly negative) amount of $\lambda_{\Omega_{k}}^R  \cdot \left(x_k - c_k\right)$. 
	\end{enumerate}
\end{enumerate}
Here, we focus on the problem of economic dispatch, and leave out the problem of unit commitment for future work. As commonly employed in electricity market computation, we assume the DC approximation of power flows \cite{Song06}. We then make a modeling assumption that the conventional generators' generation cost functions are quadratic {in its power production}. Furthermore, we do not consider generator capacity constraints in this paper, and consider them to be approximated by properly chosen quadratic functions. We note that, these modeling assumptions are not restrictive as the proposed method of finding NE can be straightforwardly generalized to cases where cost functions are higher order polynomials and generator capacities are considered. 
%

Before continuing, we summarize the notations of the relevant variables in Table \ref{nomen}. 
\begin{table}[t!]
\caption{Nomenclature\label{nomen}}
	\renewcommand{\arraystretch}{1.3}
	\begin{center}
		\begin{tabular}{@{}p{2.3cm}@{}@{}p{5.7cm}@{}}
	    	$C_i^D \left(\cdot\right), C_j^R \left(\cdot\right)$ & Cost functions of the DA conventional generator $i$ and RT conventional generator $j$.\\
            $\alpha_i^D, \beta_i^{D}$ & Quadratic and linear cost coefficients of DA generator $i$. \\
			$\alpha_j^R, \beta_j^R$ & Quadratic and linear cost coefficients of RT generator $j$. \\
			$q_i^D, q_j^R$ & Power dispatches of the DA conventional generator $i$ and RT conventional generator $j$.\\
			$\bm{q}^D, \bm{q}^R$ & Vectors of power dispatches of the DA and RT conventional generators. \\
			$\tilde{q}_u^D, \tilde{q}_u^R$ & Net nodal power injections at bus $u$ at the DA and RT markets.\\
			$\bm{\tilde{q}}^D, \bm{\tilde{q}}^R$ & Vectors of the net nodal power injections at the DA and RT markets. \\
			$L_u^D$ & Nodal load at bus $u$ at the DA market.\\
			$\bm{L}^D$ & Vector of nodal loads at the DA market.\\
			$c_k$ & Firm power commitment submitted by RPP $k$. \\
			$\bm{c}$ & Vector of firm power commitment submitted by RPPs.\\
			$\{\tilde{c}_k\}$ & Set of DA commitments that is a solution to the best response condition for the RPPs. \\
			$x_k$ & {Realization of the power output of RPP} $k$. \\
			$\bm{x}$ & {Vector of realization of power outputs of RPPs.} \\
			$\lambda_u^D, \lambda_u^R$ & LMPs at DA and RT markets for bus $u$. \\
			$\bm{\lambda}^D, \bm{\lambda}^R$ & Vectors of LMPs at DA and RT markets. \\
			$\mc{P}_k, \pi_k$ & The realized and expected payment of RPP $k$. \\
			$\bm{\pi}$ & The vector of expected payment of RPPs. \\
			$T^{(m,n)}$ & Line capacity of line between bus $m$ and $n$. \\
			$\bm{T}^{c, D}, \bm{T}^{c, R}$ & Vectors of capacities of the congested lines in the DA and RT market. \\
			$\Omega_{k}$ & Index of the bus where RPP $k$ is located.  \\
			$o$ & Index of the Slack bus. \\
			$\mc{N}, N$ & Set of buses, and number of buses. \\
			$I, J$ & Number of DA and RT conventional generators. \\
			$K$ & Number of RPPs. \\
			$n_T^{c, D}, n_T^{c, R}$ & Number of congested lines in the DA and RT market. \\
			$S_T^{c, D}, S_T^{c, R}$ & Sets of congested lines in the DA and RT markets. \\
			$S_T$ & Set of lines. \\
			$S_G^D, S_G^R$ & Sets of DA and RT conventional generators. \\
			$S_R$ & Set of RPPs. \\
			$S_{G,u}^D, S_{G,u}^R, S_{R,u}$ & Sets of DA and RT conventional generators and RPPs located on bus $u$.  \\
			$PTDF_{u,v}^{(m,n)}$ & Change in the flow of line $(m,n)$ for the injection of $1$ unit of power at bus $u$ and withdrawal of it from bus $v$. \\
			$\mbox{diag}(\cdot)$ & {If the input is a matrix, the return is its diagonal as a vector; if the input is a vector, the return is a diagonal matrix with the given vector as the diagonal.}
		\end{tabular}
	\end{center}
\end{table}

\subsection{DA and RT Market Clearing}
We now describe the details of the ISO's optimal dispatch problems that clear the DA and RT markets. 

\paragraph{DA market} In the DA market, the ISO takes the RPPs' commitments $\{c_k\}$ as firm, and then schedules the DA conventional generators to meet the net loads by solving the following DA economic dispatch problem:
\begingroup
\allowdisplaybreaks
\begin{subequations}
	\begin{align} \label{eq::DAOPF}
	& \hspace{0pt} \min_{\bm{q}^D} \hspace{3pt} \sum_{i \in S_G^D} C_i^D \left(q_i^D\right) = \sum_{i \in S_G^D} \hspace{-3pt} \left(\frac{1}{2} \alpha_i^D \cdot (q_i^D)^2 + \beta_i^{D} q_i^D\right)   \\
	& \hspace{5pt}\st\sum_{i \in S_G^{D}} q_i^D = \sum_{u \in \mc{N}} L_u^D - \sum_{k \in S_R} c_k,  \label{eq::LoadBalanceDAOPF} \\
	& \big| \sum_{u \in \mc{N}} PTDF_{u , o}^{(m,n)} \cdot \tilde{q}_{u}^D - \sum_{v \in \mc{N}} PTDF_{v , o}^{(m,n)} \cdot \tilde{q}_{v}^D \big| \nonumber \\
	& \hspace{80pt} \leq T^{(m,n)}, \hspace{7pt} \forall (m,n) \in S_T, \label{eq::DAOPFInequal}
	\end{align}
\end{subequations}
\endgroup
where PTDF refers to power transmission distribution factor, and {$\tilde{q}_u^D = \sum_{i \in S_{G,u}^D} q_i^D + \sum_{k \in S_{R,u}} c_k - L_u^D$} is the nodal net power injection at bus $u$. Note that at some buses there may be no DA conventional generator and/or no RPP.  We write the vector $\tilde{\bm{q}}^D$ as:
\begin{align}
\tilde{\bm{q}}^D = E_G^D  \bm{q}^D + E_{R}  \bm{c} -  \bm{L}^D, 
\end{align}
where the element on row $r$ and column $t$ of $E_G^D$ is $1$ if the DA conventional generator $t$ is located on bus $r$, and is $0$ otherwise. Similarly, the element on row $r$ and column $t$ of $E_{R}$ is $1$ if the $\mbox{RPP}_t$ is located on bus $r$, and is $0$ otherwise. 
From solving the DA economic dispatch on \eqref{eq::DAOPF}-\eqref{eq::DAOPFInequal}, we get the DA-LMPs $\left(\bm{\lambda}^D\right)$. The payment to RPP $k$ (located at bus $\Omega_{k}$) at the DA market is $\lambda_{\Omega_{k}}^D  c_k $.

\paragraph{RT market} In the RT market, the RPPs observe their actual power generation $\{x_i\}$. The deviations between the RPPs' DA commitments and their actual power generation is settled by optimally dispatching the RT conventional generators:
\begingroup
\allowdisplaybreaks
\begin{subequations}
	\begin{align} 
	& \hspace{0pt} \min_{\bm{q}^R} \hspace{3pt} \sum_{j \in S_G^R} C_j^R \left(\hat{q}_j^R\right) = \hspace{-5pt} \sum_{j \in S_G^R} \hspace{-3pt} \left(\frac{1}{2} \alpha_j^R \cdot (\hat{q}_j^R)^2 + \beta_j^R  \hat{q}_j^R\right) \label{eq::RTOPF}  \\
	& \hspace{5pt} \st \sum_{j \in S_G^R} q_j^R = \sum_{k \in S_R} \left(c_k - x_k\right) \notag  \\
	& \hspace{50pt} = \bm{1}^{T}  \bm{L}^R + \bm{1}^{T}  \bm{L}^D 
- \bm{1}^{T}  \bm{q}^D - \bm{1}^{T}  \bm{x},\label{eq::LoadBalanceRTOPF}\\
	& \bigl| \sum_{u \in \mc{N}} PTDF_{u , o}^{(m,n)} \cdot \tilde{q}_{u}^R - \sum_{v \in \mc{N}} PTDF_{v , o}^{(m,n)} \cdot \tilde{q}_{v}^R \bigr| \nonumber \\
	& \hspace{80pt} \leq T^{(m,n)}, \hspace{7pt} \forall (m,n) \in S_T, \label{eq::RTOPFInequal}
	\end{align}
\end{subequations}
\endgroup
where $\tilde{q}_u^R = \sum_{j \in S_{G,u}^R} q_j^R + \sum_{i \in S_{G,u}^D} q_i^D + \sum_{k \in S_{R,u}} x_k - L_u^D - L_u^R$ is the net nodal power injection at bus $u$ in the RT market. $\hat{q}_j^R = q_j^R + q_i^D$ if the RT generator $j$ participated in the DA market as DA generator $i$, otherwise, $\hat{q}_j^R = q_j^R$.
Note that on some buses there may be no RT conventional generator and$/$or no RPP. We write the vector $\tilde{\bm{q}}^R$ as:
\begin{align}
\tilde{\bm{q}}^R = E_G^R  \bm{q}^R + E_G^D  \bm{q}^D + E_{R}  \bm{x} - \bm{L}^D - \bm{L}^R,
\end{align}
{where} the element on row $r$ and column $t$ of $E_G^R$ is $1$ if the RT conventional generator $t$ is located on bus $r$, and is $0$ otherwise.
\begin{RK}
$\bm{L}^R$ is the vector of \emph{fictitious} nodal loads at the RT market, and all the components of this vector are zero. 
The reason for keeping $\bm{L}^R$ in the formulations is that it helps to derive the nodal LMPs in the RT market (cf. \eqref{eq::Change_in_RT_Dispatch} and \eqref{eq::RT_Price_Def} in Appendix \ref{proof_RT_PriceDispatch}.) 
\end{RK}
\begin{RK}
As we mentioned in Section \ref{subsec::TwoSttlmntMrkt}, there may be cases where some of the conventional generators participate in both the DA and RT markets. If generator $i$ in the DA market is the same as the generator $j$ in the RT market, then $q_i^D$ (derived from solving \eqref{eq::DAOPF}-\eqref{eq::DAOPFInequal}) is its dispatch in the DA market, and $q_j^R$ (derived from solving \eqref{eq::RTOPF}-\eqref{eq::RTOPFInequal}) is its dispatch in the RT market. Hence the total dispatch of this generator in the two-settlement market is $q_i^D + q_j^R$. 
\end{RK}

From solving RT economic dispatch \eqref{eq::RTOPF}-\eqref{eq::RTOPFInequal}, we get the RT-LMPs $\left(\bm{\lambda}^R\right)$. The payment {to $\mbox{RPP}_k$}, located at bus $\Omega_{k}$, at the RT market is $\lambda_{\Omega_{k}}^R \cdot \left(x_k - c_k\right) $.

The total realized payment to $\mbox{RPP}_k$ is 
\begin{align} \label{eq::realizedPayoff}
\mc{P}_k = \lambda_{\Omega_{k}}^D   c_k + \lambda_{\Omega_{k}}^R  \cdot \left(x_k - c_k\right). 
\end{align}
The expected total payment to $\mbox{RPP}_k$ is thus 
\begin{align} \label{eq::expectedPayoff}
\pi_k = \mathbb{E}\left[\mc{P}_k\right] = \lambda_{\Omega_{k}}^D  c_k + \mathbb{E} \left[\lambda_{\Omega_{k}}^R \cdot \left(x_k - c_k\right)\right]{.}
\end{align}

\subsection{Outcome of the mechanism: commitment game and its Nash equilibria} \label{subsec::commtmntGame}
{Now we} analyze the \emph{outcome} of the proposed mechanism (cf. Section \ref{subsec::TwoSttlmntMrkt}). This amounts to {analyzing} the {Nash} equilibria {of the two-settlement market}. In the proposed mechanism, each RPP has one decision variable, its DA commitment. Intuitively, {$\mbox{RPP}_k$ will} choose a $c_k$ that maximizes its expected payoff \eqref{eq::expectedPayoff}. The strategic behaviors of the RPPs can be modeled 
as a non-cooperative game, termed the \textit{commitment game}: 
\begin{itemize}
	\item \textit{Players}: The set of RPPs participating in the DA-RT market: $S_R = \left\{ 1,\cdots,K \right\}.$
	\item \textit{Strategies}: The {set of} firm {generation} commitments made by the RPPs. 
	\item \textit{Payoffs}: Each RPP \textit{k}'s payoff is its expected payment $\pi_k$ defined in \eqref{eq::expectedPayoff}.
\end{itemize}
The solution concept that best predicts the outcome in such a non-cooperative game is 
the Nash equilibria (NE): At a pure NE, no player has an incentive to unilaterally change its strategy, i.e., every RPP is best responding (i.e. playing an optimal commitment level) 
given all other RPPs' commitments. Specifically, for the commitment game, 

\noindent \textbf{\textit{Pure NE}:} A DA commitment profile $\left(c_1^{\star}, \cdots, c_K^{\star}\right)$ is a pure NE {if and only if}, for every {$\mbox{RPP}_k \in S_R$}, we have
\begin{align} \label{eq::def_PureNE}
\pi_k \left(c_k^{\star}, c_{-k}^{\star}\right) \geq \pi_k \left(c_k, c_{-k}^{\star}\right), \hspace{10pt} \forall c_k \in \Gamma_k,
\end{align}
where $c_{-k}^{\star}$ is the {vector of commitments} of the other RPPs except $\mbox{RPP}_k$ at the NE, and $ \Gamma_k$ is the strategy space of $\mbox{RPP}_k$. We allow each RPP to submit any real number as its DA commitment, \textit{i.e.} $\Gamma_k = \mathbb{R}.$

In order to analyze the commitment game, {it would be convenient to} have closed form expressions of the DA and RT LMPs. We show in the next section that, \textit{assuming the knowledge of which lines are congested}, a closed form expression of the corresponding LMPs can be derived.
\section{Closed form expressions of LMPs} \label{sec::closedFormLMPs}
In this section, we derive the closed form expressions of the LMPs in the DA and RT markets. 

\subsection{{DA Market LMPs}}
The DA economic dispatch problem in \eqref{eq::DAOPF}-\eqref{eq::DAOPFInequal} is a convex {quadratic} optimization problem. The KKT conditions are necessary and sufficient conditions for the {optimality of \eqref{eq::DAOPF}-\eqref{eq::DAOPFInequal}}. 
Here, our objective is to leverage the KKT conditions to obtain closed-form expressions of the LMPs. The difficulty lies in that we do not know a-priori which inequalities (corresponding to the transmission line flow constraints) in \eqref{eq::DAOPFInequal} are binding at the optimal solution. 
The key idea to overcome this difficulty is the following: If we \emph{assume} the knowledge of which lines are congested, which we term \textit{assumed DA congestion pattern}, the rest of the KKT conditions can then be solved in closed form. We summarize this result as follows.
\begin{theorem} \label{Theorem_DA_PriceDispatch}
For an assumed DA congestion pattern in the DA market, the optimal solution of the DA economic dispatch in \eqref{eq::DAOPF}-\eqref{eq::DAOPFInequal} is a linear function of the DA commitments of the RPPs {as}
\begin{align} \label{eq::q_DA_equation}
\bm{q}^D = G_1^D  \bm{c} + G_2^D.
\end{align}
{Similarly}, the DA-LMPs at the DA market is a linear function of the DA commitments of the RPPs {as}
\begin{align} \label{eq::DALMP}
\bm{\lambda}^D = H_1^D  \bm{c} + H_2^D.
\end{align}
\end{theorem}
The proof of Theorem \ref{Theorem_DA_PriceDispatch} and the closed forms of matrices $G_1^D$, $G_2^D$, $H_1^D$, and $H_2^D$ can be found in Appendix \ref{proof_DA_PriceDispatch}.
\vspace{-2mm}
\subsection{{RT market LMPs}}
The approach for finding closed form expressions of RT market LMPs is similar to that for the DA LMPs. \emph{Assuming the knowledge of the congestion pattern at the RT market}, the rest of the KKT conditions for the RT dispatch problem \eqref{eq::RTOPF}-\eqref{eq::RTOPFInequal} can be solved in closed form.
\begin{theorem} \label{Theorem_RT_PriceDispatch}
For an assumed RT congestion pattern in the RT market, a given set of power dispatches of DA conventional generators in the DA market, the optimal solution of the RT economic dispatch in \eqref{eq::RTOPF}-\eqref{eq::RTOPFInequal} is a linear function of the RPPs' DA commitments and RT  realizations {as}
\begin{align} \label{eq::q_RT_equation}
\bm{q}^R = G_1^R  \bm{c} + G_2^R  \bm{x} + G_3^R.
\end{align}
{Similarly,} the RT-LMPs 
is a linear function of the RPPs' DA commitments and RT realizations {as}
\begin{align} \label{eq::RTLMP}
\bm{\lambda}^R = H_1^R  \bm{c} + H_2^R  \bm{x} + H_3^R.
\end{align}
\end{theorem}
The proof of Theorem \ref{Theorem_RT_PriceDispatch} and the closed forms of matrices $G_1^R$, $G_2^R$, $G_3^R$, $H_1^R$, $H_2^R$, and $H_3^R$ can be found in Appendix \ref{proof_RT_PriceDispatch}.
%

%
\section{Algorithm for Finding Nash Equilibria}
The key idea that helps finding an NE among the RPPs (cf. \eqref{eq::def_PureNE}) is that, if the DA and RT congestion patterns at an NE are known, finding the NE $\{c_k^\star\}$ given the congestion patterns is easy. Then, one can search and test if any DA and RT congestion patterns are indeed the ones that lead to an NE.  

\subsection{Best responses assuming DA and RT congestion patterns} 
Assuming any pair of DA and RT congestion patterns, we have from the last section the corresponding expressions of the LMPs $\bm{\lambda}^D$ and $\bm{\lambda}^R$ \eqref{eq::DALMP} and \eqref{eq::RTLMP}. 
Substituting these LMP expressions in \eqref{eq::expectedPayoff}, 
the vector of payoffs of the RPPs {$\bm{\pi} \triangleq \left[\pi_1, \cdots, \pi_K \right]^{\top}$} becomes 
\begin{align} \label{eq::Exp_payoff_vec}
\bm{\pi}&= \mbox{diag} \left( (E_R)^{\top}  \bm{\lambda}^D \right)  \bm{c}  + \mathbb{E} \left[ \mbox{diag} \left( (E_R)^{\top}  \bm{\lambda}^R \right)  \left(\bm{x} - \bm{c} \right) \right].
\end{align}
%
%
From the linearity of the LMP expressions in \eqref{eq::DALMP} and \eqref{eq::RTLMP}, the payoffs of the RPPs in \eqref{eq::Exp_payoff_vec} are \emph{concave quadratic functions}. As such, the NE condition in \eqref{eq::def_PureNE} reduces to {the following} set of \emph{linear equations} 
\begin{align} \label{eq::BestResponse}
\frac{d \pi_k}{d c_k} \Big|_{\left(c_1,\cdots,c_K\right) =  \left(c_1^{\star}, \cdots, c_K^{\star}\right)} = 0, \hspace{20pt} \forall k \in S_R{.}
\end{align} 
The details of this set of linear best response equations are provided below: 
\begin{align} 
    &\left[ \frac{d \pi_k}{d c_k} \right] \triangleq \left[\frac{d \pi_1}{d c_1}, \cdots, \frac{d \pi_K}{d c_K} \right]^{\top} \nonumber \\
    &= \big(\mbox{diag} \left( \mbox{diag} \left((E_R)^{\top}  \left( H_1^D - H_1^R \right) \right) \right) \nonumber \\
    & \hspace{40pt} + (E_R)^{\top}\left( H_1^D - H_1^R \right) \big) \bm{c} \nonumber \\
    & \hspace{10pt} +  (E_R)^{\top}  \left( H_2^D - H_2^R \bm{\mu} - H_3^R \right) \nonumber \\
    & \hspace{40pt} + \mbox{diag} \left( \mbox{diag} \left( (E_R)^{\top}  H_1^R \right) \right)  \bm{\mu} = 0.\label{eq::DerivPayoffsVector}
\end{align}

The solution to this set of linear equations \eqref{eq::BestResponse}-\eqref{eq::DerivPayoffsVector}, denoted by $\{\tilde{c}_k\}$, 
provides an \emph{NE candidate} of the commitment game. Whether this renders a \emph{true NE} depends on if the assumed DA and RT congestion patterns are indeed the ones at a true NE. Notably, computing such an NE candidate is {as easy as} solving a set of linear equations.

\subsection{Algorithm to Find Pure NE}
\subsubsection{Consistency of DA congestion pattern} \label{sec::FindingNE}
For an assumed DA congestion pattern to be one at a true NE, it is necessary that the corresponding NE candidate $\{\tilde{c}_k\}$ satisfies the following condition: Given $\{\tilde{c}_k\}$ as the DA firm commitments from the RPPs, as the ISO solves the optimal DA dispatch problem (cf. \eqref{eq::DAOPF}-\eqref{eq::DAOPFInequal}), the \emph{resulting actual DA congestion pattern is the same as the assumed one}.  
Otherwise, the assumed DA congestion pattern is an incorrect guess of that at a true NE. 


%
\subsubsection{Probability of consistency of RT congestion pattern} \label{sec:RTcons}
Similarly, we can check the consistency between the assumed and actual RT congestion patterns. This is however more subtle than checking the consistency for DA congestion patterns. Note that, the RPPs' decisions, firm commitments $\{c_k\}$, are made at DA, when their actual generations at RT are still \emph{uncertain}. As such, at DA, the future \emph{actual} RT congestion pattern when the ISO clears the RT market is uncertain. To be precise, given a set of RPPs' DA commitments $\{c_k\}$ and the ISO's DA dispatch decisions, the optimal RT dispatch depends on the uncertain generation $\{x_k\}$, and there is \emph{a probability distribution} over what RT congestion pattern would result from the optimal RT dispatch \cite{Tong17}. 

In this work, we proceed with an approximation of the above situation. Instead of having each RPP {to} consider a probability distribution over RT congestion patterns, we assume that each RPP just {considers} one RT congestion pattern. We will then evaluate the \emph{probability of this one congestion pattern appearing at RT}: When this probability is sufficiently high, we argue that this approximation is a close one. 

Now, with the above approximation, instead of checking the consistency between the assumed and actual congestion patterns as in DA, we check the \emph{probability of such consistency}. Specifically, given an \emph{assumed RT congestion pattern}, the corresponding NE candidate $\{\tilde{c}_k\}$, and the resulting DA optimal dispatch, a) depending on the realized generation $\{x_k\}$, the ISO would solve an optimal RT dispatch problem \eqref{eq::RTOPF}-\eqref{eq::RTOPFInequal}, resulting in an RT congestion pattern, and b) based on the uncertainty in $\{x_k\}$, \emph{the probability of this resulting RT congestion pattern being the same as the assumed one is computed as the probability of consistency.} 

Accordingly, a) we first require the ``absolute'' consistency of DA congestion pattern as in Section \ref{sec::FindingNE}, (otherwise the NE candidate is not a true NE for sure), and then b) the probability of consistency of RT congestion pattern can be interpreted as the \emph{probability that this NE candidate is a true NE}. 


\subsubsection{The proposed algorithm of finding NE}
Based on the above development, we provide the algorithm for finding the pure NE of the commitment game in Algorithm \ref{Algorithm}. In this algorithm, various heuristics can be employed in searching over congestion patterns. One approach is to collect a set of possible DA and RT congestion patterns and simply cycle through all of them. As straightforward as this may sound, it can actually be quite effective in practice, especially because the set of possible congestion patterns are often reasonably limited in power networks \cite{Tong17, Javadi17, Roald18, Roald19}. 

\begin{algorithm}
\begin{algorithmic}[1]
\caption{ Searching For Pure NE} \label{Algorithm}
 \renewcommand{\algorithmicrequire}{\textbf{Input:}}
 \renewcommand{\algorithmicensure}{\textbf{Output:}}
 \REQUIRE Set of candidate pairs of DA and RT congestion patterns
 \ENSURE  Set of pure NE with probabilities of being valid
 \\ \textit{Initialization} : Set of pure NE = empty set
  \FOR {each candidate pair of DA and RT congestion patterns:}
  \STATE Determine the linear functions for the DA and RT LMPs for these \textit{assumed} DA and RT congestion patterns as in \eqref{eq::DALMP} and \eqref{eq::RTLMP}.
  \STATE Compute the DA commitments for this NE candidate, $\{\tilde{c}_k\}$, as in \eqref{eq::BestResponse}.
  \STATE Applying the DA commitments, solve the DA economic dispatch problem \eqref{eq::DAOPF}-\eqref{eq::DAOPFInequal}, and find the \textit{actual} DA congestion pattern. 
  \IF {the \textit{actual} DA congestion pattern is the same as the \textit{assumed} DA congestion pattern}
  \STATE a) Based on the probability distribution over the RPPs' generation $\{x_k\}$, compute the \emph{probability} that the actual congestion pattern from solving the RT optimal dispatch problem is the same as the assumed RT congestion pattern. 
  \STATE b) Add the DA commitments $\{\tilde{c}_k\}$ to the set of pure NE, and record the above probability of it being valid. 
  \ENDIF
  \ENDFOR
 \RETURN Set of pure NE with probabilities of being valid. 
 \end{algorithmic} 
 \end{algorithm}
\vspace{-2mm}
\section{Simulations}
In this section, we demonstrate the main results of the paper with simulations on the IEEE 14-bus system. 
A Python module is written for simulating the two-settlement market and the proposed algorithm for finding NE. The module and simulation codes are available at \cite{PythonModule}. 
The nodal demands, nominal parameters of RPPs (which will be further varied), and parameters of conventional generators are listed in Tables \ref{tbl::LoadData},  \ref{tbl::RPPData}, and \ref{tbl::ConvGenData}. 
We note that, the standard deviation of RPPs' generation in Table \ref{tbl::RPPData} can be interpreted as the standard deviation of the DA forecast error of the renewable generation. In this nominal case, the forecast error has a standard deviation of $15\%$ of the point forecast of generation. This is in fact quite a conservative assumption on forecast accuracy, as it is less accurate than a typical DA forecast (e.g., with a $11.9\%$ std/mean as reported in \cite{hodge2012wind}). 

In what follows, we will first present results on finding the NE among two RPPs located at two different buses. We will then show that the NE converges to social efficiency as the number of RPPs grows. 
%
%
\begin{table}[t]
	\caption{Data of Nodal Demands \vspace{-3mm}}
	\label{tbl::LoadData}
	\begin{center}
		\begin{tabular}{|p{4mm}|p{2.5mm}|p{2.5mm}|p{2.5mm}|p{2.5mm}|p{2.5mm}|p{2.5mm}|p{2.5mm}|}
			\hline
			bus & 0 & 1 & 2 & 3 & 4 & 5 & 6 \\
			\Xhline{0.5\arrayrulewidth}
			load & 0 & 43.4 & 64.8 & 41.6 & 15.2 & 22.4 & 20   \\ \Xhline{2\arrayrulewidth}
			bus & 7 & 8 & 9 & 10 & 11 & 12 & 13 \\ \Xhline{0.5\arrayrulewidth}
			load & 50 & 59 & 18 & 27 & 32.2 & 27.6 & 29.8 \\ \hline
		\end{tabular}
	\end{center}
\end{table}
\begin{table}[t]
	\caption{Data of RPPs \vspace{-3mm}}
	\label{tbl::RPPData}
	\renewcommand{\arraystretch}{1.1}
	\begin{center}
		\begin{tabular}{|c|c|c|}
			\hline
			bus & Mean (MWh) & Standard Deviation (MWh) \\
			\hline
			4 & 70 & 10.5   \\ \hline
			11 & 50 & 7.5  \\ \hline
		\end{tabular}
	\end{center}
\end{table}
\begin{table}[t]
\addtolength{\tabcolsep}{-1.8pt}
\caption{Data of DA and RT Generators \vspace{-3mm}}
\label{tbl::ConvGenData}
\begin{center}
\begin{tabular}{|c|p{9mm}|p{9mm}|c|p{9mm}|p{9mm}|}
\hline
\multicolumn{3}{|c|}{\centering DA Conventional Generators} & \multicolumn{3}{c|}{\centering RT Conventional Generators}  \\
\hline 
\centering bus & 
\parbox{6mm}{\begin{align*} & \hspace{7pt} \alpha^D \\ & \hspace{-6pt} ( \frac{\$}{\mbox{(MWh)}^2} )  \end{align*}} & 
\parbox{6mm}{\begin{align*} & \hspace{8pt} \beta^D \\ & \hspace{0pt} ( \frac{\$}{\mbox{MWh}} ) \end{align*}} & 
\centering bus & 
\parbox{6mm}{\begin{align*} & \hspace{7pt} \alpha^R \\ & \hspace{-6pt} ( \frac{\$}{\mbox{(MWh)}^2} )  \end{align*}} & 
\parbox{6mm}{\begin{align*} & \hspace{8pt} \beta^R \\ & \hspace{0pt} ( \frac{\$}{\mbox{MWh}} ) \end{align*}} 
\\ [-1ex] \hline
 7    &  0.06        & 3.51        & 4    & 0.24         & 9.35   \\
8    &  0.09        & 3.89        & 12   & 0.26         & 11.51  \\
11   &  0.08        & 2.15        &      &              &       \\
\hline
\end{tabular}
\end{center}
\end{table}
%
%
\begin{figure*}[htp]
\centering
\begin{subfigure}{.67\columnwidth}
\includegraphics[width=\columnwidth]{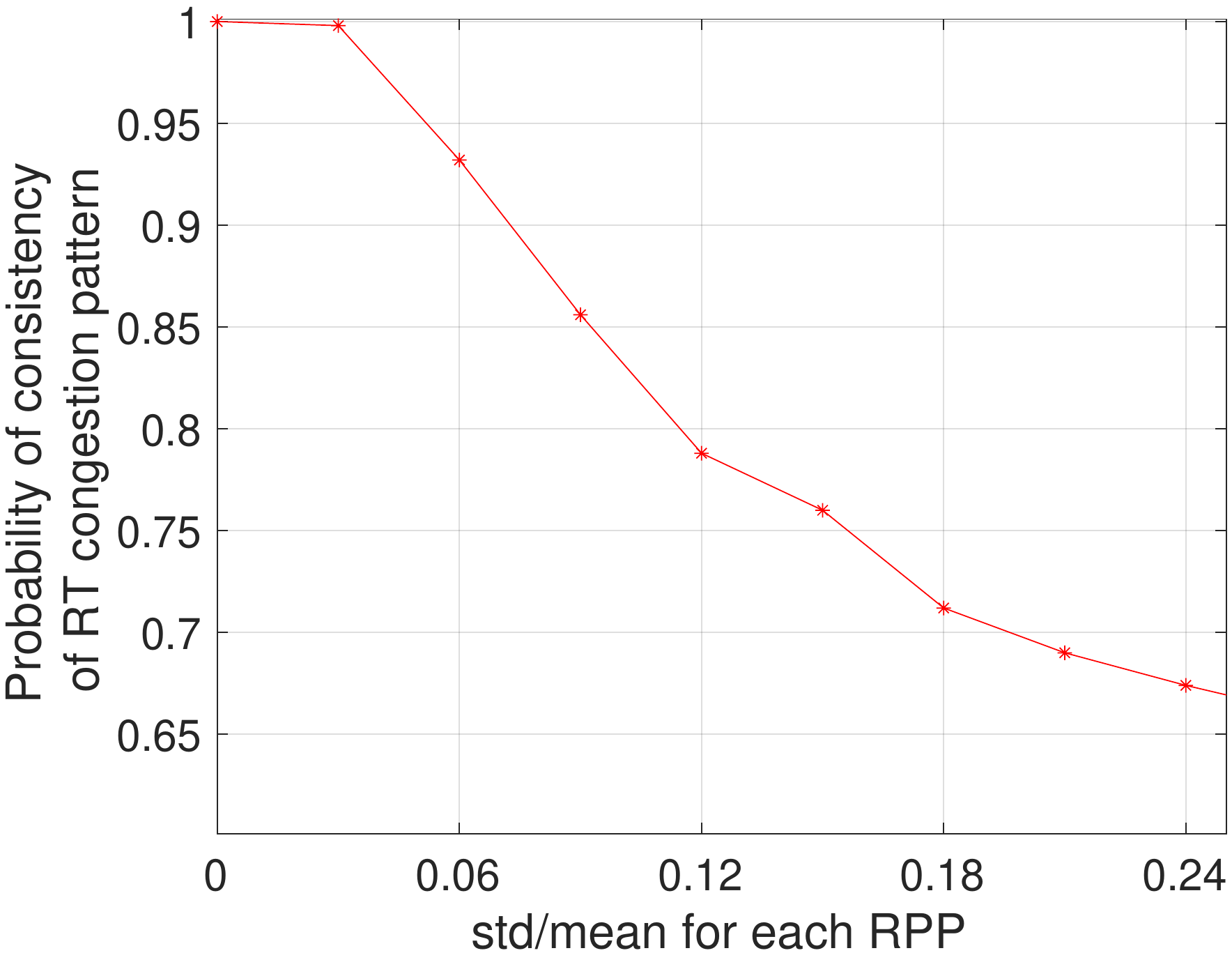}%
\caption{}%
\label{fig::probNE}%
\end{subfigure}\hfill%
\begin{subfigure}{.67\columnwidth}
\includegraphics[width=\columnwidth]{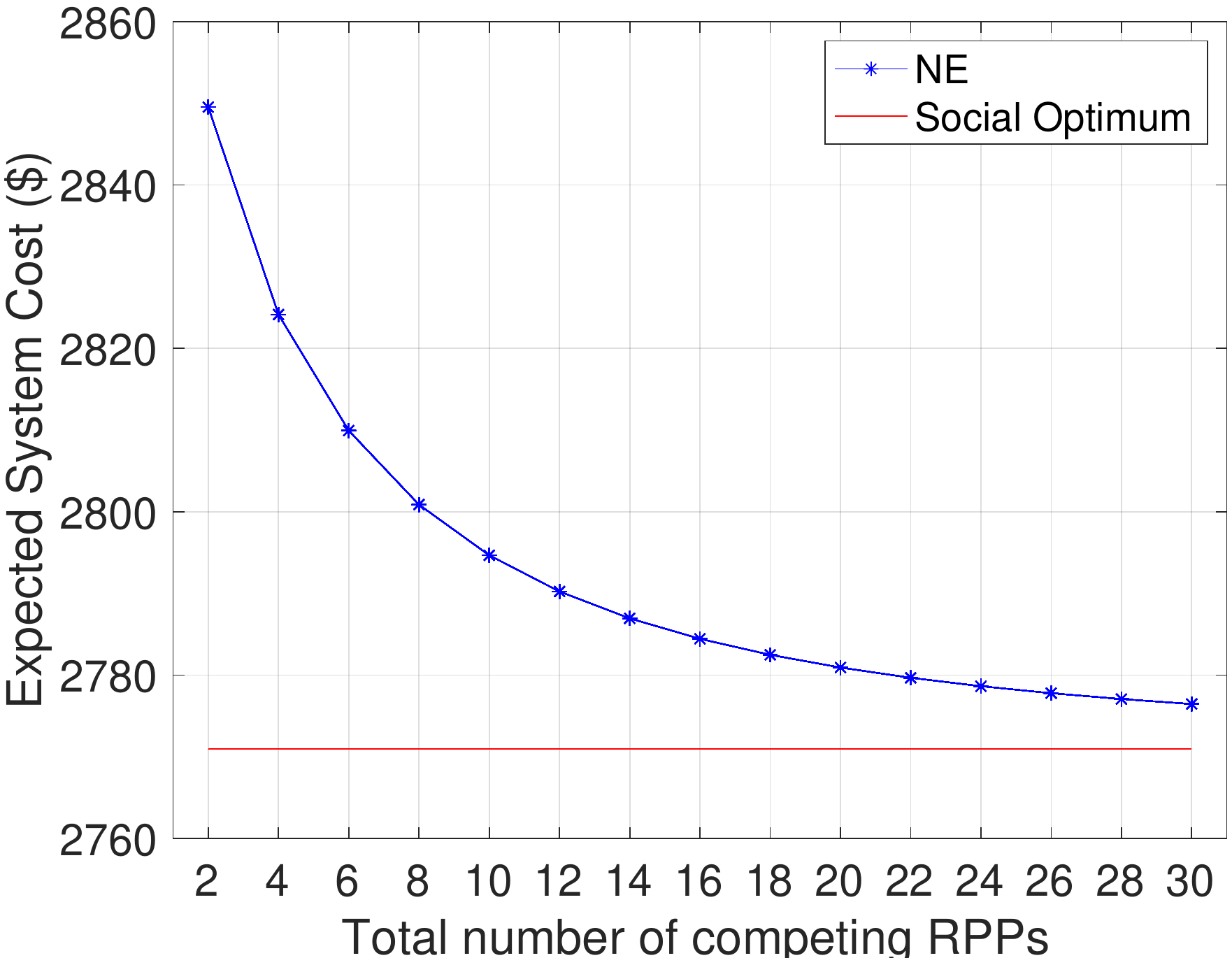}%
\caption{}%
\label{fig::SysCost}%
\end{subfigure}%
\hspace{3pt}
\begin{subfigure}{.67\columnwidth}
\includegraphics[width=\columnwidth]{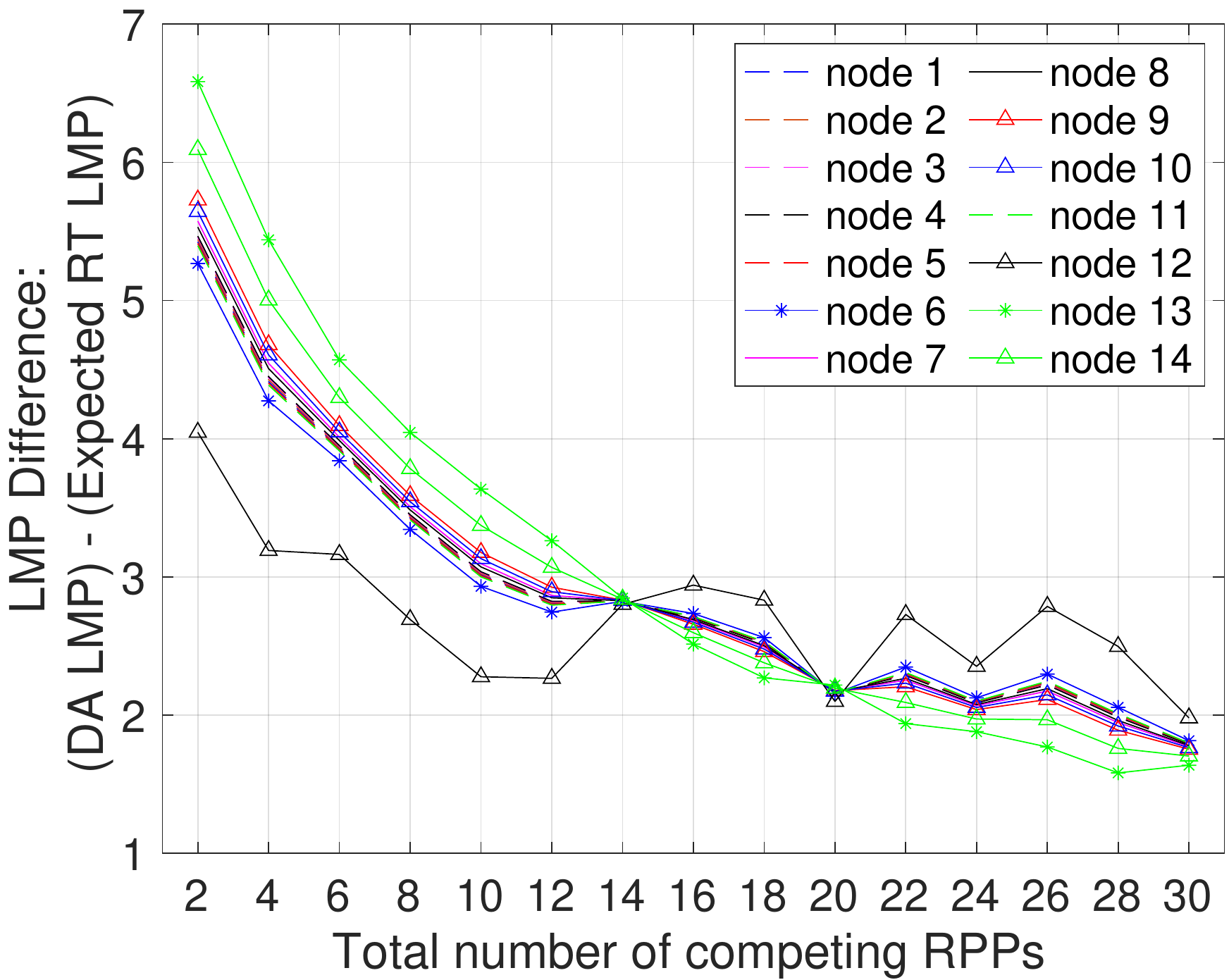}%
\caption{}%
\label{fig::LMPDiff}%
\end{subfigure}%
\caption{(a) Probability of consistency of RT congestion pattern vs. the uncertainty of RPPs, (b) Expected system cost: competing RPPs vs. social optimum as the number of RPPs grows, (c) Difference between DA and expected RT LMPs as the number of RPPs grows.}
\label{figabc}
\end{figure*}
%
%


\vspace{-2mm}
\subsection{Finding pure NE}
In our simulations, it suffices to search for NE over congestion patterns assuming a) the DA and RT congestion patterns are the same, and b) no more than $2$ lines are simultaneously congested. The following pure NE is found: $\left(c_1^{\star} , c_2^{\star}\right) = \left(77.270 , 46.095\right)$.
The corresponding DA congestion pattern is that a single line $\#19$ is congested. 

Next, we check the probability of consistency of the RT congestion pattern (cf. Section \ref{sec:RTcons}) which is assumed to be the same as the DA one. We employ an Monte Carlo approach for computing this probability. We generate $500$ scenarios for the renewable generation (cf. Table \ref{tbl::RPPData}) assuming normal distribution. For each scenario, we clear the RT market by optimally dispatching the RT conventional generators (cf. \eqref{eq::RTOPF} - \eqref{eq::RTOPFInequal}) and find the \textit{actual} RT congestion pattern. The probability of consistency of RT congestion pattern is then computed as the ratio between a) the number of scenarios where the assumed and actual RT congestion patterns agree, and b) the total number of scenarios (in our case $500$). The resulting probability is $76\%$ for the nominal case in Table \ref{tbl::RPPData}, a reasonably high consistency. 

%

Intuitively, this probability of consistency at RT depends on the level of uncertainty of the renewables. As such, we evaluate this probability with varying level of uncertainty: For std/mean of RPPs' generation ranging from $0$ to $25\%$, we repeat the scenario based Monte Carlo computation of the probability as above, and plot the resulting probabilities of consistency at RT in Figure \ref{fig::probNE}. 
%
As expected, when there is no uncertainty, the assumed RT congestion pattern appears with $100\%$ probability. Even with a $25\%$ std/mean of the RPPs' generation (corresponding to very poor forecast), the probability of consistency at RT is still above $65\%$. 
\vspace{-2mm}
\subsection{Convergence of NE to social efficiency}
We now investigate the important question of how close the NE is from social efficiency. Inspired by the intuition from \cite{ZhaoPES18}, we expect that the gap between the NE and the social optimum decreases as the number of RPPs grows. 
Here, we break up the RPP at each of the two buses into an increassing number of equal-sized market participants. 
For each case, we recompute the NE. 
The expected system costs for all these cases are plotted in Fig. \ref{fig::SysCost}, and compared with the social optimum obtained by solving a two-stage stochastic optimization problem. 
%
Indeed, the expected system cost at the NE decreases as the total number of RPPs grows, and converges to that at the social optimum, (although we only plotted for up to a total of $30$ RPPs, and the convergence is numerically confirmed as the number of RPPs further increases). Details for computing the  social optimum can be found in Appendix \ref{Appn::SocialOptimum}, where a penalty factor 
of $\psi=5000$ is employed (cf. \eqref{eq::SO_OPF} - \eqref{eq::SO_OPFInequal_22}). We further plot the trends of the differences between the DA and RT LMPs as the number of RPPs grows in Figure \ref{fig::LMPDiff}. Notably, a) due to the congestion at the NE, the LMPs at all the buses are different; nonetheless, b) their DA-RT differences all decrease as the NE converges to social efficiency. 

As a result, we observe that the following advantage of the proposed mechanism for RPPs: Even with the ISO fully relying on RPPs' DA commitments and only solving deterministic economic dispatch, the NE of the proposed market mechanism still converges to the social optimum as if a full-blown two-stage stochastic optimization is solved. The intuition is that, enabled by the proposed mechanism, the competition among the RPPs successfully pushes the NE toward social efficiency, again without the ISO considering any of their uncertainty whatsoever. 

%
\vspace{-3mm}
\section{Conclusion} \label{sec:conc}
We have proposed a simple market mechanism for integrating renewable power producers (RPPs) in power systems. In it, a) the RPPs submit firm power delivery commitments in the DA market, b) the ISO solves a deterministic DA economic dispatch problem, and the RPPs are paid according to the resulting DA LMPs and their DA commitments, c) in the RT market, the RPPs' generation are realized, d) the ISO solves a deterministic RT economic dispatch problem, and the RPPs are paid/charged according to the resulting RT LMPs and their realizations' deviations from the DA commitments. We developed a novel method for efficiently finding pure Nash {equilibria} (NE) of the market among the RPPs. {In particular}, we search over congestion patterns, compute NE candidates given assumed congestion patterns, and verify if they are true NE or not. Simulation results show that, the NE of the proposed market mechanism converges to social efficiency as the number of the RPPs grows. Notably, this is achieved without having the ISO to consider any uncertainty whatsoever in its dispatch decisions. 

\bibliographystyle{IEEEtran}
{\bibliography{TPS}}

\begin{thebibliography}{10}
\providecommand{\url}[1]{#1}
\csname url@samestyle\endcsname
\providecommand{\newblock}{\relax}
\providecommand{\bibinfo}[2]{#2}
\providecommand{\BIBentrySTDinterwordspacing}{\spaceskip=0pt\relax}
\providecommand{\BIBentryALTinterwordstretchfactor}{4}
\providecommand{\BIBentryALTinterwordspacing}{\spaceskip=\fontdimen2\font plus
\BIBentryALTinterwordstretchfactor\fontdimen3\font minus
  \fontdimen4\font\relax}
\providecommand{\BIBforeignlanguage}[2]{{%
\expandafter\ifx\csname l@#1\endcsname\relax
\typeout{** WARNING: IEEEtran.bst: No hyphenation pattern has been}%
\typeout{** loaded for the language `#1'. Using the pattern for}%
\typeout{** the default language instead.}%
\else
\language=\csname l@#1\endcsname
\fi
#2}}
\providecommand{\BIBdecl}{\relax}
\BIBdecl

\bibitem{Pinson2013}
P.~Pinson, ``Wind energy: Forecasting challenges for its operational
  management,'' \emph{Statistical Science}, vol.~28, no.~4, pp. 564--585, 2013.

\bibitem{ConejoBook}
A.~J. Conejo, M.~Carri\'on, and J.~M. Morales, \emph{Decision making under
  uncertainty in electricity markets}.\hskip 1em plus 0.5em minus 0.4em\relax
  Springer, 2010.

\bibitem{shapiro2009lectures}
A.~Shapiro, D.~Dentcheva \emph{et~al.}, \emph{Lectures on stochastic
  programming: modeling and theory}.\hskip 1em plus 0.5em minus 0.4em\relax
  SIAM, 2009, vol.~9.

\bibitem{Hobbs2012}
C.~De~Jonghe, B.~Hobbs, and R.~Belmans, ``Optimal generation mix with
  short-term demand response and wind penetration,'' \emph{IEEE Transactions on
  Power Systems}, vol.~27, no.~2, pp. 830--839, May 2012.

\bibitem{ZouSDDiP2019}
J.~Zou, S.~Ahmed, and X.~A. Sun, ``Stochastic dual dynamic integer
  programming,'' \emph{Mathematical Programming}, vol. 175, no. 1-2, pp.
  461--502, 2019.

\bibitem{ZouUC2018}
------, ``Multistage stochastic unit commitment using stochastic dual dynamic
  integer programming,'' \emph{IEEE Transactions on Power Systems}, vol.~34,
  no.~3, pp. 1814--1823, 2018.

\bibitem{Hu07}
X.~{Hu} and D.~{Ralph}, ``Using {EPECs} to model bilevel games in restructured
  electricity markets with locational prices,'' \emph{Operations Research},
  vol.~55, no.~5, pp. 809--827, 2007.

\bibitem{anderson2002using}
E.~J. Anderson and A.~B. Philpott, ``Using supply functions for offering
  generation into an electricity market,'' \emph{Operations Research}, vol.~50,
  no.~3, pp. 477--489, 2002.

\bibitem{anderson2008finding}
E.~J. Anderson and X.~Hu, ``Finding supply function equilibria with asymmetric
  firms,'' \emph{Operations Research}, vol.~56, no.~3, pp. 697--711, 2008.

\bibitem{johari2011}
R.~Johari and J.~N. Tsitsiklis, ``Parameterized supply function bidding:
  Equilibrium and efficiency,'' \emph{Operations research}, vol.~59, no.~5, pp.
  1079--1089, 2011.

\bibitem{NaLi15}
N.~Li, L.~Chen, and M.~A. Dahleh, ``Demand response using linear supply
  function bidding,'' \emph{IEEE Transactions on Smart Grid}, vol.~6, no.~4,
  pp. 1827--1838, July 2015.

\bibitem{Xu16}
Y.~Xu, N.~Li, and S.~H. Low, ``Demand response with capacity constrained supply
  function bidding,'' \emph{IEEE Transactions on Power Systems}, vol.~31,
  no.~2, pp. 1377--1394, March 2016.

\bibitem{lin2017structural}
W.~Lin and E.~Bitar, ``A structural characterization of market power in power
  markets,'' \emph{arXiv:1709.09302}, 2017.

\bibitem{allaz1993cournot}
B.~Allaz and J.-L. Vila, ``Cournot competition, forward markets and
  efficiency,'' \emph{Journal of Economic theory}, vol.~59, no.~1, pp. 1--16,
  1993.

\bibitem{yao2008modeling}
J.~Yao, I.~Adler, and S.~S. Oren, ``Modeling and computing two-settlement
  oligopolistic equilibrium in a congested electricity network,''
  \emph{Operations Research}, vol.~56, no.~1, pp. 34--47, 2008.

\bibitem{cai2017role}
D.~Cai, S.~Bose, and A.~Wierman, ``On the role of a market maker in networked
  cournot competition,'' \emph{arXiv preprint arXiv:1701.08896}, 2017.

\bibitem{Nayyar13}
A.~Nayyar, K.~Poolla, and P.~Varaiya, ``A statistically robust payment sharing
  mechanism for an aggregate of renewable energy producers,'' \emph{Proc.
  European Control Conference (ECC)}, pp. 3025--3031, 2013.

\bibitem{Bitar11}
E.~{Baeyens}, E.~Y. {Bitar}, P.~P. {Khargonekar}, and K.~{Poolla}, ``Wind
  energy aggregation: A coalitional game approach,'' in \emph{2011 50th IEEE
  Conference on Decision and Control and European Control Conference}, Dec
  2011, pp. 3000--3007.

\bibitem{Harirchi14}
F.~Harirchi, T.~Vincent, and D.~Yang, ``Optimal payment sharing mechanism for
  renewable energy aggregation,'' in \emph{Proc. 2014 IEEE International
  Conference on Smart Grid Communications (SmartGridComm)}, Nov 2014, pp.
  608--613.

\bibitem{Zhao2016}
Y.~Zhao and H.~Khazaei, ``An incentive compatible profit allocation mechanism
  for renewable energy aggregation,'' in \emph{Proc. 2016 IEEE Power Energy
  Society General Meeting}, July 2016.

\bibitem{Zhao19}
H.~{Khazaei} and Y.~{Zhao}, ``Indirect mechanism design for efficient and
  stable renewable energy aggregation,'' \emph{IEEE Transactions on Power
  Systems}, vol.~34, no.~2, pp. 1033--1042, March 2019.

\bibitem{ZhaoPES18}
------, ``Competitive market with renewable power producers achieves asymptotic
  social efficiency,'' in \emph{2018 IEEE Power Energy Society General Meeting
  (PESGM)}, Aug 2018.

\bibitem{Tong17}
Y.~{Ji}, R.~J. {Thomas}, and L.~{Tong}, ``Probabilistic forecasting of
  real-time lmp and network congestion,'' \emph{IEEE Transactions on Power
  Systems}, vol.~32, no.~2, pp. 831--841, March 2017.

\bibitem{Javadi17}
M.~{Javadi}, M.~{Hong}, R.~N. {Angarita}, S.~H. {Hosseini}, and J.~N. {Jiang},
  ``Identification of simultaneously congested transmission lines in power
  systems operation and market analysis,'' \emph{IEEE Transactions on Power
  Systems}, vol.~32, no.~3, pp. 1772--1781, May 2017.

\bibitem{Roald18}
Y.~{Ng}, S.~{Misra}, L.~A. {Roald}, and S.~{Backhaus}, ``Statistical learning
  for dc optimal power flow,'' in \emph{2018 Power Systems Computation
  Conference (PSCC)}, June 2018, pp. 1--7.

\bibitem{Roald19}
S.~Misra, L.~Roald, and Y.~Ng, ``Learning for constrained optimization:
  identifying optimal active constraint sets,'' \emph{arXiv preprint
  arXiv:1802.09639}, 2018.

\bibitem{Song06}
Y.-H. Song and X.-F. Wang, \emph{Operation of Market-Oriented Power
  Systems}.\hskip 1em plus 0.5em minus 0.4em\relax Springer, 2006.

\bibitem{PythonModule}
\BIBentryALTinterwordspacing
H.~Khazaei. (2019) Python module: Market-for-renewables. [Online]. Available:
  \url{https://github.com/hoseinkh/Market-for-Renewables/releases/latest}
\BIBentrySTDinterwordspacing

\bibitem{hodge2012wind}
B.-M. Hodge \emph{et~al.}, ``Wind power forecasting error distributions: An
  international comparison,'' in \emph{11th International Workshop on
  Large-Scale Integration of Wind Power into Power Systems as well as on
  Transmission Networks for Offshore Wind Power Plants}, 2012.

\end{thebibliography}

\newpage
\appendices
\section{Proof of Theorem \ref{Theorem_DA_PriceDispatch}} \label{proof_DA_PriceDispatch}
Let $S_T^{c, D}$ be the set of the lines that are congested in the DA market, for which each line $(m,n)$ in it is congested in the direction of bus $m$ to bus $n$. First, let's rewrite \eqref{eq::DAOPFInequal} for those lines where the inequality is binding:
\begin{align}
A^D  \tilde{\bm{q}}^{D} = \bm{T}^{c, D}, 
\end{align}
where $A^D$ and $\bm{T}^{c, D}$ are derived from writing \eqref{eq::DAOPFInequal} only for binding inequalities.
There is one more set of matrices that we need to define:
\begin{align}
& A_G^D \triangleq A^D  E_G^D  \nn \\
& A_R^D \triangleq A^D E_{R},
\end{align}
\begin{align}
    \Upsilon^D \triangleq  \left[ \begin{matrix}
                        \alpha_1^D & & 0 \\
                        \hspace{1pt}& \ddots & \hspace{1pt} \\
                        0 & \hspace{1pt} & \alpha_I^D
                    \end{matrix} \right]
\end{align}
and
\begin{align}
Z^D \triangleq 
\left[
\begin{array}{c|c|c}
\Upsilon^D & \left(A_G^D\right)^{\top} & \bm{1}_{I \times 1} \vspace{2pt} 
\dividerline{0.5ex}{2.3ex}
A_G^D & \bm{0}_{n_T^{c,D} \times n_T^{c,D}} & \bm{0}_{n_T^{c,D} \times 1} 
\dividerline{1.5ex}{2.3ex}
\bm{1}_{1 \times I} & \bm{0}_{1 \times n_T^{c,D}} & 0
\end{array}
\right]
\end{align}
Writing the KKT optimality conditions for \eqref{eq::DAOPF}-\eqref{eq::DAOPFInequal}, we have:
\begin{align} \label{eq::KKT_DA_eqtns}
\hspace{-3pt}
Z^D \hspace{-1pt} \cdot
\hspace{-1pt}
\left[ \hspace{2pt} 
\begin{array}{c}
\begin{matrix}
\begin{matrix}
\hspace{8pt} 
\bm{q}^D \\
\hspace{1pt}
\end{matrix} \\[-8pt]
\hline \\[-20pt]
\begin{matrix}
\hspace{1pt} \\
\hspace{8pt}
\bm{\gamma}^D \\
\hspace{1pt}
\end{matrix}
\\[-8pt]  \hline \\[-8pt]
\hspace{2pt}
\tau^D
\end{matrix}
\end{array}
\hspace{-2pt} \right]
\hspace{-2pt}
=
\hspace{-2pt}
\left[ \hspace{0pt} 
\begin{array}{c}
\begin{matrix}
\begin{matrix}
\hspace{0pt}  \\[-8pt]
- \bm{\beta}^D \\[-8pt]
\hspace{0pt}
\end{matrix} \\
\hline \\[-8pt]
\rule{0pt}{0ex} A^D  \bm{L}^D + \bm{T}^{c, D} 
\rule{0pt}{0ex} - A_R^D  \bm{c} \\[2pt]
\hline \\[-8pt]
\bm{1}^{\top}  \bm{L}^D - \bm{1}^{\top}  \bm{c}
\end{matrix}
\end{array}
\hspace{-4pt}
\right]
\end{align}
Where $\bm{\beta}^D = \left[ \beta_1^D , \cdots , \beta_I^D \right]^{\top}$, and $\bm{\gamma}^D = \left[ \gamma_{1}^D , \cdots , \gamma_{n_T^{c,D}}^D \right]^{\top}$, where $\gamma_t^D$ is the dual variable of the congested line $t$ in the DA economic dispatch problem (cf. \eqref{eq::DAOPFInequal}). $\tau^D$ is the dual variable of the power balance equation in the DA economic dispatch (cf. \eqref{eq::LoadBalanceDAOPF}). 
The first $I$ equations in \eqref{eq::KKT_DA_eqtns} is for the \textit{stationarity} conditions. The following $n_T^{c,D}$ equations corresponds to the \textit{complementary slackness conditions} for those inequalities that are bindings in \eqref{eq::DAOPFInequal} by setting the line flow of those lines equal to their capacities. The last equation in \eqref{eq::KKT_DA_eqtns} is the \textit{primal feasibility} condition corresponding the load balance equation in \eqref{eq::LoadBalanceDAOPF}. 

Let define $W^D$ be the the first $I$ rows of the inverse of the $Z^D$, \textit{i.e.}
\begin{align}
W^D \triangleq \left[ \bm{\mathcal{I}}_{I \times I} \ \bigg| \ \bm{0}_{I \times (n_T^{c,D} + 1)}  \right]    \left( Z^D \right)^{-1}.
\end{align}
Where $\bm{\mathcal{I}}_{I \times I}$ is the $I \times I$ identity matrix.
By solving \eqref{eq::KKT_DA_eqtns} we have:
\begin{align} \label{eq::Close_form_q_DA}
    \bm{q}^D = W^D 
\hspace{0pt}
\cdot \left[ \hspace{0pt} 
\begin{array}{c}
\begin{matrix}
\begin{matrix}
\hspace{0pt}  \\[-8pt]
\hspace{4pt} - \bm{\beta}^D \\[-8pt]
\hspace{0pt}
\end{matrix} \\
\hline \\[-8pt]
\rule{0pt}{0ex} A^D  \bm{L}^D + \bm{T}^{c, D} - A_R^D  \bm{c} \\[2pt]
\hline \\[-8pt]
\bm{1}^{\top}  \bm{L}^D - \bm{1}^{\top}  \bm{c}
\end{matrix}
\end{array}
\hspace{-4pt}
\right]
\end{align}
Now we rewrite the closed form formula of the DA dispatches of the DA conventional generators derived in \eqref{eq::Close_form_q_DA}:
\begin{align} \label{eq::Simple_Close_form_q_DA}
\bm{q}^D = G_1^D  \bm{c} + G_2^D
\end{align}
Where $G_1^D$ and $G_2^D$ are derived from \eqref{eq::Close_form_q_DA}. 

One approach to calculate the DA-LMPs is to perform a sensitivity analysis of the changes of the $\bm{q}^D$ w.r.t. the changes in the nodal demands, and then calculate the the changes in the system cost (cf. objective function in \eqref{eq::DAOPF}).
In order to calculate the DA-LMP at bus $u$, we hypothetically change the demand at this bus for a small amount of power, \textit{i.e.} $\epsilon$ units of power, and calculate the the change in the DA system cost. Let's define the vector of new nodal demands after changing demand at bus $u$ as $\bm{L}_u^{D,new} = \bm{L}^D + \epsilon \cdot \mbox{\textbf{\^{\mdseries I}}}_u$, where $\mbox{\textbf{\^{\mdseries I}}}_u$ is the unit vector of size $N$ whose $u^{th}$ element is 1 and other elements are zero. 
First, we derive the vector of \textit{new} power dispatches of the DA conventional generators for the new demand profile $\bm{L}_u^{D,new}$.
From \eqref{eq::Close_form_q_DA} we have:
\begin{align} \label{eq::Change_in_DA_Dispatch}
    \bm{q}^D\Big|_{\bm{L}_u^{D,new}} - \bm{q}^D \Big|_{\bm{L}^D} = \ \epsilon \cdot W^D 
\hspace{0pt}
\left[ \hspace{0pt} 
\begin{array}{c}
\begin{matrix}
\begin{matrix}
\hspace{0pt}  \\[-8pt]
\hspace{4pt} \bm{0} \\[-8pt]
\hspace{0pt}
\end{matrix} \\
\hline \\[-8pt]
\rule{0pt}{0ex} A_{(u)}^D \\[2pt]
\hline \\[-8pt]
1
\end{matrix}
\end{array}
\hspace{-4pt}
\right]
\end{align}
Where $A_{(u)}^D$ is the $u^{\mbox{th}}$ column of $A^D$.
Note that, unless otherwise explicitly said, $\bm{q}^D$ refers to the set of original power dispatches of the DA conventional generators for clearing the market, with no small change in the load profile. \\
Now, we can calculate the limit of ratio of change in the system cost (corresponding to the change of $\epsilon$ unit of load at bus $u$) to $\epsilon$, which is DA-LMP at this bus:
\begin{align} \label{eq::DA_Price_Def}
    \lambda_u^D = \lim_{\epsilon \rightarrow 0}  \dfrac{\splitfrac{ \big( \sum_{i \in S_G^D} C_i^D \left(q_i^D\right) \Big|_{q_i^D \in \bm{q}^D\big|_{\bm{L}_u^{D,new}}}}{ - \sum_{i \in S_G^D} C_i^D \left(q_i^D\right) \Big|_{q_i^D \in \bm{q}^D\big|_{\bm{L}^D}}\big)}}{\epsilon}
\end{align}
From \eqref{eq::Change_in_DA_Dispatch} and \eqref{eq::DA_Price_Def} we can get the following closed form formula for the DA price at bus $u$:
\begin{align}
    \lambda_u^D = \left( \Upsilon^D \bm{q}^D +  \bm{\beta}^D  \right)^{\top}  W^D 
    \left[ \hspace{0pt} 
\begin{array}{c}
\begin{matrix}
\begin{matrix}
\hspace{0pt}  \\[-8pt]
\hspace{4pt} \bm{0} \\[-8pt]
\hspace{0pt}
\end{matrix} \\
\hline \\[-8pt]
\rule{0pt}{0ex} A_{(u)}^D \\[2pt]
\hline \\[-8pt]
1
\end{matrix}
\end{array}
\hspace{-4pt}
\right]
\end{align}
The vector of DA-LMPs is:
\begin{align} \label{eq::DA_LMPs_detailed}
    \bm{\lambda}^D = \left[ \bm{0}_{N \times I} \big| \left( A^D \right)^{\top} \big| \bm{1}_{N \times 1} \right] \left( W^D \right)^{\top} \left(  \Upsilon^D    \bm{q}^D  + \bm{\beta}^D    \right)
\end{align}
%
%
%
%
Replacing $\bm{q}^D$ from \eqref{eq::Simple_Close_form_q_DA} and with a little algebra we have:
\begin{align} \label{eq::Simple_Close_form_DA_LMP}
\bm{\lambda}^D = H_1^D  \bm{c} + H_2^D
\end{align}
Where $H_1^D$ and $H_2^D$ are derived from \eqref{eq::DA_LMPs_detailed}.
\section{Proof of Theorem \ref{Theorem_RT_PriceDispatch}}  \label{proof_RT_PriceDispatch}
Assume that the set of congested lines in the RT market is $S_T^{c, R}$, for which each line $(m,n)$ in it is congested in the direction of bus $m$ to bus $n$. We can rewrite the inequalities in \eqref{eq::RTOPFInequal} as equalities for the binding ones, and ignore the rest of them, as following:
\begin{align} \label{eq::EqualLineCons}
A^R  \tilde{\bm{q}}^R = \bm{T}^{c, R}
\end{align}
Where $A^R$ is derived from writing \eqref{eq::RTOPFInequal} \textit{only} for binding inequalities. Also we need to define the following matrices:
\begin{align}
& A_G^R \triangleq A^R  E_G^R, \nonumber \\
& A_R^R \triangleq A^R  E_R, 
\end{align}
\begin{align}
    \Upsilon^R \triangleq  \left[ \begin{matrix}
                        \alpha_1^R & & 0 \\
                        \hspace{1pt}& \ddots & \hspace{1pt} \\
                        0 & \hspace{1pt} & \alpha_J^R
                    \end{matrix} \right],
\end{align}
and
\begin{align} \label{eq::Z_RT}
Z^R \triangleq 
\left[
\begin{array}{c|c|c}
\Upsilon^R & \left(A_G^R\right)^{\top} & \bm{1}_{J \times 1} \vspace{2pt} 
\dividerline{0.5ex}{2.3ex}
A_G^R & \bm{0}_{n_T^{c,R} \times n_T^{c,R}} & \bm{0}_{n_T^{c,R} \times 1} 
\dividerline{1.5ex}{2.3ex}
\bm{1}_{1 \times J} & \bm{0}_{1 \times n_T^{c,R}} & 0
\end{array}
\right].
\end{align}
%
%
%
Writing the KKT optimality conditions, we have:
\begin{flalign} \label{eq::KKT_RT_eqtns}
\hspace{-6pt}
Z^R
\hspace{-1pt}
\cdot
\hspace{-1pt}
\left[ \hspace{0pt} 
\begin{array}{c}
\begin{matrix}
\begin{matrix}
\hspace{5pt} 
\bm{q}^R
\hspace{-8pt}
\\
\hspace{-8pt}
\end{matrix} \\[-8pt]
\hline 
\\[-20pt]
\begin{matrix}
\hspace{-8pt} \\
\hspace{5pt}
\bm{\gamma}^R \hspace{-8pt} \\
\hspace{-8pt}
\end{matrix}
\\[-8pt] 
\hline 
\\[-8pt]
\hspace{2pt}
\tau^R \hspace{-10pt}
\end{matrix}
\end{array}
\hspace{-5pt} \right] 
\hspace{-2pt}
= 
\hspace{-2pt}
\left[ \hspace{1pt} 
\begin{array}{c}
\begin{matrix}
\begin{matrix}
\hspace{3pt} \\[-7pt]
- \bm{\beta}^R
- E_G^{DR}  \Upsilon^D  \bm{q}^{D}  \\[-7pt]
\hspace{3pt}
\end{matrix} \\
\hline \\[-7pt]
\rule{0pt}{0ex} A^R  \bm{L}^R  - A^R E_G^D  \bm{q}^D  \\[1pt]
\rule{0pt}{0ex} + A^R  \bm{L}^D + \bm{T}^{c, R} - A_R^R  \bm{x}  \\[3pt]
\hline \\[-7pt]
\bm{1}^{\top}  \bm{L}^R + \bm{1}^{\top}  \bm{L}^D  \\[1pt]
- \bm{1}^{\top}  \bm{q}^D - \bm{1}^{\top}  \bm{x}
\end{matrix}
\end{array}
\hspace{-3pt}
\right]
\end{flalign}
Where $\bm{\beta}^R = \left[ \beta_i^R , \cdots , \beta_J^R \right]^{\top}$, and $\bm{\gamma}^R = \left[ \gamma_{1}^R , \cdots , \gamma_{n_T^{c,R}}^R \right]^{\top}$, where $\gamma_t^R$ is the dual variable of the congested line $t$ in the RT economic dispatch problem (cf. \eqref{eq::RTOPFInequal}). $\tau^R$ is the dual variable of the power balance equation in the RT economic dispatch (cf. \eqref{eq::LoadBalanceRTOPF}). $E_G^{DR}$ is a $J \times I$ matrix, whose element on row $r$ and column $t$ is one if the RT conventional generator $r$ also participates in the DA market as the DA conventional generator $t$, otherwise, it is zero. The first $J$ equations in \eqref{eq::KKT_RT_eqtns} is for the \textit{stationarity} conditions. The following $n_T^{c,R}$ equations corresponds to the \textit{complementary slackness conditions} for those inequalities that are bindings in \eqref{eq::RTOPFInequal} by setting the line flow of those lines equal to their capacities.
The last equation in \eqref{eq::KKT_RT_eqtns} is the \textit{primal feasibility} condition corresponding the load balance equation in \eqref{eq::LoadBalanceRTOPF}.

Let us define $W^R$ be the the first $J$ rows of the inverse of the $Z^R$, \textit{i.e.}
\begin{align}
W^R \triangleq \left[ \bm{\mathcal{I}}_{J \times J} \ \bigg| \ \bm{0}_{J \times (n_T^{c,R} + 1)}  \right]    \left( Z^R \right)^{-1}.
\end{align}
By solving \eqref{eq::KKT_DA_eqtns} we have:
\begingroup
\allowdisplaybreaks
\begin{align} \label{eq::Close_form_q_RT}
    \bm{q}^R = W^R \cdot
\hspace{0pt}
\left[ \hspace{0pt} 
\begin{array}{c}
\begin{matrix}
\begin{matrix}
\hspace{0pt}  \\[-8pt]
- \bm{\beta}^R   - E_G^{DR}  \Upsilon^D  \bm{q}^D \\[-8pt]
\hspace{0pt}
\end{matrix} \\
\hline \\[-8pt]
\rule{0pt}{0ex} A^R  \bm{L}^R + A^R  \bm{L}^D   \nn
\\ - A^R E_G^D  \bm{q}^D + \bm{T}^{c, R} - A_R^R  \bm{x} \\[2pt]
\hline \\[-8pt]
\bm{1}^{\top}  \bm{L}^R + \bm{1}^{\top}  \bm{L}^D  - \bm{1}^{\top}  \bm{q}^D - \bm{1}^{\top}  \bm{x}
\end{matrix}
\end{array}
\hspace{-4pt}
\right] \\
\end{align}
\endgroup
Now we can rewrite the closed form formula of the RT dispatches of the RT conventional generators derived in \eqref{eq::Close_form_q_RT} using the closed form formula for $\bm{q}^D$ derived in \eqref{eq::q_DA_equation}:
\begin{align} \label{eq::Simple_Close_form_q_RT}
\bm{q}^R = G_1^R  \bm{c} + G_2^R  \bm{x} + G_3^R
\end{align}
Where $G_1^R$, $G_2^R$ and $G_3^R$ are derived from \eqref{eq::Close_form_q_RT}. 
\\
%
To calculate the RT-LMP at bus $u$, we need to hypothetically change the RT demand at this bus for small amount of power, \textit{i.e.} $\epsilon$ units of power, and calculate the the change in the RT system cost. First, we need to calculate the vector of \textit{new} power dispatches of the RT conventional generators for the change of $\epsilon$ units of RT demand at bus $u$. Let's define $\bm{L}_u^{R,new} = \bm{L}^R + \epsilon \cdot \mbox{\textbf{\^{\mdseries I}}}_u$, where $\mbox{\textbf{\^{\mdseries I}}}_u$ is the unit vector of size $N$ whose $u^{{th}}$ element is 1 and other elements are zero. From \eqref{eq::Close_form_q_RT} we have:
\begin{align} \label{eq::Change_in_RT_Dispatch}
    \bm{q}^R\Big|_{\bm{L}_u^{R,new}} - \bm{q}^R \Big|_{\bm{L}^R} = \ \epsilon \cdot W^R 
\hspace{0pt}
\left[ \hspace{0pt} 
\begin{array}{c}
\begin{matrix}
\begin{matrix}
\hspace{0pt}  \\[-8pt]
\hspace{4pt} \bm{0} \\[-8pt]
\hspace{0pt}
\end{matrix} \\
\hline \\[-8pt]
\rule{0pt}{0ex} A_{(u)}^R \\[2pt]
\hline \\[-8pt]
1
\end{matrix}
\end{array}
\hspace{-4pt}
\right]
\end{align}
Note that, unless otherwise explicitly said, $\bm{q}^R$ refers to the vector of original power dispatches of the RT conventional generators for clearing the RT market with no small change in the load profile. \\
Now, we can calculate the limit of ratio of change in the RT system cost (corresponding to the change of $\epsilon$ unit of RT demand at bus $u$) to $\epsilon$, which is RT-LMP at this bus:
\begin{align} \label{eq::RT_Price_Def}
    \lambda_u^R = \lim_{\epsilon \rightarrow 0}  \dfrac{\splitfrac{ \big( \sum_{j \in S_G^R} C_j^R \left(\hat{q}_j^R\right) \Big|_{q_j^R \in \bm{q}^R\big|_{\bm{L}_u^{R,new}}}}{ - \sum_{j \in S_G^R} C_j^R \left(\hat{q}_j^R\right) \Big|_{q_j^R \in \bm{q}^R\big|_{\bm{L}^R}}\big)}}{\epsilon}
\end{align}
From \eqref{eq::Change_in_RT_Dispatch} and \eqref{eq::RT_Price_Def} we can get the following closed form formula for the RT price at bus $u$:
\begin{align}
     \hspace{-10pt} \lambda_u^R = \hspace{0pt} \left( \Upsilon^R  \bm{q}^R + E_G^{DR}  \Upsilon^D  \bm{q}^D   + \bm{\beta}^R \right)^{\top}
     W^R
\hspace{0pt}
\left[ \hspace{0pt} 
\begin{array}{c}
\begin{matrix}
\begin{matrix}
\hspace{0pt}  \\[-8pt]
\hspace{4pt} \bm{0} \\[-8pt]
\hspace{0pt}
\end{matrix} \\
\hline \\[-8pt]
\rule{0pt}{0ex} A_{(u)}^R \\[2pt]
\hline \\[-8pt]
1
\end{matrix}
\end{array}
\hspace{-4pt}
\right] 
\end{align}
The vector of the RT-LMPs is
\begin{align} \label{eq::RT_LMPs_detailed}
    \bm{\lambda}^R &=\left[ \bm{0}_{N \times J} \big| \left( A^R \right)^{\top} \big| \bm{1}_{N \times 1} \right] \left( W^R \right)^{\top} \big( \Upsilon^R  \bm{q}^R \nn \\
    & \hspace{75pt} + E_G^{DR}  \Upsilon^D  \bm{q}^D   + \bm{\beta}^R \big)
\end{align}
%
%
%
Using the closed formulas for $\bm{q}^D$ and $\bm{q}^R$ in \eqref{eq::Simple_Close_form_q_DA} and \eqref{eq::Simple_Close_form_q_RT} and with a little algebra we get the following closed form formula for the $\bm{\lambda}^R$:
\begin{align}
\bm{\lambda}^R = H_1^R  \bm{c} + H_2^R  \bm{x} + H_3^R
\end{align}
where $H_1^R$, $H_2^R$ and $H_3^R$ are derived from \eqref{eq::RT_LMPs_detailed}.
\section{Calculation of Social Optimum} \label{Appn::SocialOptimum}
The social optimum refers to the case where the ISO knows all the information of the RPPs as well as DA and RT conventional generators and minimizes the expected system cost. In designing market mechanisms the primary goal is to achieve social efficiency, meaning that the expected system cost under the mechanism would be equal (or at least close) to the expected system cost at the social optimum. We shall emphasize that when calculating the social optimum, we \textit{assume} that ISO knows everything, including the joint probability distribution of the RPPs.

In solving the social optimum problem, we generate a set of possible scenarios for the uncertain variables, \textit{i.e.} power generation of the RPPs. Each scenario has a probability, and ISO may generate some scenarios with higher probabilities and some other scenarios with lower probabilities. Based on the the generated scenarios, ISO solves the following optimization problem
\begingroup
\allowdisplaybreaks
\begin{subequations}
	\begin{align} 
	& \hspace{-10pt} \min_{\substack{\bm{q}^D , \{ \bm{q}^{R,y}, \\ \bm{u}^{+,y}, \bm{u}^{-,y} \}} } \hspace{3pt} \sum_{i \in S_G^D} \hspace{-2pt} C_i^D \left(q_i^D\right)  + \hspace{-5pt} \sum_{y \in S_S} \zeta_y \cdot \Big\{\hspace{-3pt}  \sum_{j \in S_G^{R}} \hspace{-2pt} C_j^R \left(\hat{q}_j^{R,y}\right)  \nonumber \\
	& \hspace{50pt} + \psi \cdot \sum_{y \in S_S} \sum_{(m,n) \in S_T} \left( ( u_{(m,n)}^{+,y} + u_{(m,n)}^{-,y} )^2 \right) \Big\} \nonumber \\
	&\hspace{10pt} = \hspace{-5pt} \sum_{i \in S_G^D} \hspace{-3pt} \left(\frac{1}{2} \alpha_i^D \cdot (q_i^D)^2 + \beta_i^{D}  q_i^D\right)  + \hspace{-5pt} \sum_{y \in S_S} \zeta_y \cdot  \Big\{ \nonumber \\
	& \hspace{57pt} \sum_{j \in S_G^R} \hspace{-3pt} \left(\frac{1}{2} \alpha_j^R \cdot (\hat{q}_j^{R,y})^2 + \beta_j^R  \hat{q}_j^{R,y}\right)  \nonumber \\
	& \hspace{27pt} + \psi \cdot \sum_{y \in S_S} \sum_{(m,n) \in S_T} \left( (u_{(m,n)}^{+,y} + u_{(m,n)}^{-,y} )^2 \right) \Big\} \label{eq::SO_OPF} \\
	& \hspace{5pt} s.t. \nonumber  \\
	& \hspace{0pt} \sum_{i \in S_G^D} q_i^D + \sum_{j \in S_G^{R}} q_j^{R,y} + \sum_{k \in S_R} x_k^y = \sum_{n \in \mc{N}} L_n^D, \forall y \in S_S \label{eq::LoadBalanceSO_OPF}  \\
	& u_{(m,n)}^{+,y} \geq \hspace{-2pt} \sum_{u \in \mc{N}} PTDF_{u , o}^{(m,n)} \cdot \tilde{q}_{u}^y - \sum_{v \in \mc{N}} PTDF_{v , o}^{(m,n)} \cdot \tilde{q}_{v}^y  \nonumber \\
	& \hspace{60pt} - T^{(m,n)}, \hspace{7pt} \forall (m,n) \in S_T, \forall y \in S_S \label{eq::SO_OPFInequal_11} \\
	& u_{(m,n)}^{+,y} \geq 0   \label{eq::SO_OPFInequal_12} \\
	& u_{(m,n)}^{-,y} \geq \hspace{-2pt} \sum_{u \in \mc{N}} PTDF_{u , o}^{(n,m)} \cdot \tilde{q}_{u}^y - \sum_{v \in \mc{N}} PTDF_{v , o}^{(n,m)} \cdot \tilde{q}_{v}^y  \nonumber \\
	& \hspace{60pt} - T^{(m,n)}, \hspace{7pt} \forall (m,n) \in S_T, \forall y \in S_S \label{eq::SO_OPFInequal_21} \\
	& u_{(m,n)}^{-,y} \geq 0   \label{eq::SO_OPFInequal_22}
	\end{align}
\end{subequations}
\endgroup
where $\tilde{q}_u^y \triangleq \sum_{j \in S_{G,u}^R} q_j^{R,y} + \sum_{i \in S_{G,u}^D} q_i^D + \sum_{k \in S_{R,u}} x_k - L_u^D$ is the net power injected at bus $u$ under scenario $y$. Also, $\hat{q}_j^{R,y} = q_j^{R,y} + q_i^D$ if the RT generator $j$ participated in the DA market as DA generator $i$, otherwise, $\hat{q}_j^{R,y} = q_j^{R,y}$.
Set $S_S$ is the set of scenarios and $\zeta_y$ is the probability of scenario $y$. Note that, unlike \eqref{eq::DAOPFInequal} in DA-OPF or \eqref{eq::RTOPFInequal} in RT-OPF, here we do not impose \textit{strict} transmission line constraints because there always could exist a scenario that makes the optimization problem infeasible. Instead we impose a penalty coefficient (\textit{i.e.} $\psi$) to penalize any line flow that is greater than the transmission line capacity. The auxiliary variable $u_{(m,n)}^{+,y}$ ($u_{(m,n)}^{-,y}$) is used to represent the excess line flow of line $(m,n)$ from bus $m$ to $n$ (from bus $n$ to $m$) under scenario $y$. If the line flow of a line $(m,n)$ under scenario $y$ is greater than the capacity of that line, then one of the following two cases happens:
\begin{itemize}
    \item The line flow is in the direction of the line that is defined in the $S_T$, \textit{i.e.} from bus $m$ to bus $n$. In this case the auxiliary variable $u_{(m,n)}^{+,y}$ is equal to the absolute value of the excess line flow of line $(m,n)$.
    \item Or the line flow is in the reverse of the direction of the line that is defined in the $S_T$, \textit{i.e.} from bus $n$ to bus $m$. In this case the auxiliary variable $u_{(m,n)}^{-,y}$ is equal to the absolute value of the excess line flow of line $(m,n)$.
\end{itemize}
Note that the optimization formulation penalizes such cases. Usually $\psi$ is a large number.
\ifCLASSOPTIONcaptionsoff
  \newpage
\fi

\end{document}